\documentclass[]{article}   

\usepackage{graphicx}
\usepackage{subfigure}
\usepackage{amsmath,bm}
\usepackage{color}
\usepackage{gensymb}
\usepackage{amsfonts}
\usepackage[group-separator={,}, group-minimum-digits=4]{siunitx}
\usepackage{authblk}
\usepackage{times}
\usepackage[left=1in, right=0.7in, bottom=0.6in, top=1in]{geometry}

\title{On critical dynamics and thermodynamic efficiency\\of urban transformations}

\author[1,2,*]{Emanuele Crosato}
\author[1]{Ramil Nigmatullin}
\author[1]{Mikhail Prokopenko}
\affil[1]{Complex Systems Research Group and Centre for Complex Systems, Faculty of Engineering and IT, The University of Sydney, Sydney, NSW 2006, Australia.}
\affil[2]{CSIRO Data61, PO Box 76, Epping, NSW 1710, Australia.}
\affil[*]{emanuele.crosato@sydney.edu.au}

\date{}

\begin{document}

\maketitle

\begin{abstract}
Urban transformations within large and growing metropolitan areas often generate critical dynamics affecting social interactions, transport connectivity and income flow distribution.
We develop a statistical-mechanical model of urban transformations, exemplified for Greater Sydney, 
and derive a thermodynamic description highlighting critical regimes.
We consider urban dynamics at two time scales: fast dynamics for the distribution of population and income, modelled via the maximum entropy principle, and slower dynamics evolving the urban structure under spatially distributed competition.
We identify phase transitions between dispersed and polycentric phases, induced by varying the \emph{social disposition}---a factor balancing the suburbs' attractiveness---in contrast with the \emph{travel impedance}.
Using the Fisher information we identify critical thresholds and quantify the thermodynamic cost of urban transformation, as the minimal work required to vary the underlying parameter.
Finally, we introduce the notion of \emph{thermodynamic efficiency of urban transformation}, as the ratio of the order gained during a change to the amount of required work, showing that this measure is maximised at criticality.\\

\end{abstract}



\section{Introduction}

A city is quintessentially a complex system consisting of multiple interacting agents such as individual residents, employment centres and transport infrastructure~\cite{batty2008size, batty2013new}.
The complexity manifests itself through diverse spatial organisations: monocentric cities where most of the economic activity takes place at the CBD~\cite{alonso1964location}, polycentric cities with multiple subcentres (or ``edge'' cities)~\cite{hartshorn1989suburban, scott1990technopoles, garreau2011edge} and dispersed sprawl (or ``edgeless'') cities~\cite{lang2003edgeless}.
Moreover, cities can undergo transitions in their urban structures.
Driving such transitions are changes in the factors determining economies and diseconomies of agglomeration for both firms and residents~\cite{odland1978conditions, fujita1982multiple, richardson1995economies, louf2013modeling}.

While urban dynamics have been extensively studied in the past, a unified framework centered on thermodynamics of urban transformations has not been yet developed (see a review~\cite{bouchard2013crises}).
In particular, the analysis and modelling of urban transformations as phase transitions, defined in a rigorous thermodynamic setting, remains an open challenge, despite recent attempts in spatial economics over short time scales~\cite{louf2013modeling}.
This paper aims to refocus the research field on Urban Thermodynamics, considering critical phenomena including phase transitions in a principled way, based on the underlying thermodynamic concepts (energy potentials, entropy, order parameters, etc.), for both equilibrium and nonequilibrium scenarios.
This approach will enable systematic calibrations of such thermodynamic models with real-world data and scenarios at different overlapping time scales.

We develop a statistical-mechanical model displaying phase transitions, using the maximum entropy principle in a dynamic setting, and define the thermodynamic efficiency of urban transformations.
The model is calibrated to Greater Sydney Census data and is shown to exhibit a phase transition between a monocentric dispersed and polycentric clustered urban forms.
This phase transition is induced by the variation of the attractiveness of the residential neighbourhoods, measured by the density of local services, given the transportation cost.
While quantitative studies of urban transformation typically focus on statistical analysis of spatial evolution of cities~\cite{griffith1981evaluating, pfister2000polycentricity, mcmillen2003number, lee2007edge, louail2015uncovering}, the thermodynamic approach developed in this paper enables a rigorous analysis of critical dynamics in a wide class of urban systems, as well as quantitative explorations of diverse ``what-if'' scenarios with respect to a generic and precise efficiency measure.

Our model is based on the Boltzmann-Lotka-Volterra (BLV) method~\cite{wilson1978equilibrium, wilson2008boltzmann, wilson2011phase, osawa2017harris}.
The BLV models involve two components: a fast equilibration, ``Boltzmann'', component and a slow dynamic, ``Lotka-Volterra'', component.
The Boltzmann component applies maximum entropy principle to derive the static flow patterns of commodities and residents consistent with given spatial distributions~\cite{wilson2011entropy}.
The Lotka-Volterra component evolves the spatial distribution and the flow pattern of a commodity according to generalised Lotka-Volterra equations for spatially distributed competitors.
In our model of Greater Sydney the Lotka-Volterra equations make suburbs compete for local services, and suburbs with more services become more attractive residential places.
The resultant urban dynamics exhibit critical regimes, interpreted as urban phase transitions, where a small variation in suitably chosen (control) parameters changes the global outcomes measured via specific aggregated quantities (order parameters).

The maximum entropy method~\cite{jaynes1957information} has been applied to a variety of collective phenomena~\cite{bialek2012statistical, tkacik2015thermodynamics} and urban modelling~\cite{wilson2011entropy}, suggesting a formal analogy between urban and thermodynamic systems~\cite{wilson2009thermodynamics, morphet2013thunen, hernando2012workings}.
In studying transformations in the Greater Sydney region as thermodynamic phenomena, we construct the corresponding phase diagram with respect to suitably chosen control parameters.
In doing so we use the Fisher information, which measures the sensitivity of a probability distribution to the change in the control parameter, and diverges at critical points~\cite{brody1995geometrical, prokopenko2011relating, wang2011fisher, prokopenko2015information, crosato2018thermodynamics}.

Our analysis further deepens the analogy between urban science and thermodynamics, utilising a clear thermodynamic interpretation of the Fisher information as the second derivative of free entropy.
Specifically, we investigate the minimum work required to vary a control parameter and trace configuration entropy and internal energy, according with the first law of thermodynamics.
Crucially, the thermodynamic work is defined via Fisher information and thus can be computed solely based on probability distributions estimated from available data.
Finally, we introduce the concept of \emph{thermodynamic efficiency of urban transformation} as the ratio of the order gained during a change to the required work, and demonstrate that it is maximised at criticality for our case study.


\section{Material and methods}

\subsection{Overview of the model}

In our model, the population commutes between home and work place.
The number of people commuting between employment areas $i$ and residence areas $j$ is given by the \emph{travel-to-work} matrix $T_{ij}$.
Commuting trips have an associated cost $C_{ij}$, e.g., travelling expenses, time or distance.
$C_{ij}$ represents the structure of the transport network, which may include the roads as well as different types of public transport.
Employment areas are characterised by the average \emph{income} $I_i$ earned by the employees that, in combination with the travel-to-work matrix, provides the \emph{flow of income} $Y_{ij} = T_{ij}I_i$.
Residence areas are instead characterised by the average \emph{rent} $R_j$, and the amount of \emph{services} $S_j$ (the data used in modelling Greater Sydney is described in Supplemental Material, Sec. 1).

We develop a BLV model for the predicted income flow ${\cal{Y}}_{ij}$ in contrast with the actual income flow $Y_{ij}$ obtained from the Census.
The number of jobs available in each employment area is assumed to remain fixed, and therefore the income flowing out of each area is also fixed: $Y^{out}_i = \sum_jY_{ij}$.
On the contrary, the population is allowed to redistribute among the suburbs.
The services $S_j$ and the population $P_j$ determine the \emph{attractiveness} $A_j$ of a suburb, which defines people's preference to live in, and therefore bring their income to $j$.
When deciding where to settle, people consider the utility of living in attractive suburbs as well as the cost of commuting to work.
In our model this tradeoff is controlled by two parameters, $\alpha$ and $\gamma$, which define, respectively, how much value is attributed to suburbs with respect to their attractiveness (\emph{social disposition}) and how much discomfort is attributed to commuting trips with respect to costs (\emph{travel impedance}).

The model further allows the urban services $S_j$ (and therefore the attractiveness $A_j$) to evolve, with these dynamics being slower than the resettling of people.
When ${\cal{Y}}_{ij}$ units of income are moved from employment areas $i$ to residence areas $j$, part of it is spent on the rent $R_j$ while the remainder can be spent on the services in $j$.
Lotka-Volterra dynamics make suburbs compete for the services: if the income that can be spent in a suburb is higher than the cost of running the services in that suburb, these services of will grow, otherwise they will decrease.
Every time $S_j$ is updated, ${\cal{Y}}_{ij}$ is recomputed using the maximum entropy principle, until an equilibrium is reached such that the income spent on services matches their running cost.
This results in a converging sequence of income flow matrices from an initial ${\cal{Y}}^0_{ij}$ to a final ${\cal{Y}}^*_{ij}$.

\subsection{The Boltzmann component}

The Boltzmann component of the model, informed by the maximum entropy principle, determines the least biased flow-of-income matrix ${\cal{Y}}_{ij}$ which satisfies the constraints on the income that employment areas can produce, the attractiveness of the residence areas and the cost of travelling.
Such flow of income is the one that maximises the entropy
\begin{equation}
\label{eq:entropy-max-ent}
H({\cal{Y}}_{ij}) = -\sum_i\sum_j {\cal{Y}}_{ij} \log {\cal{Y}}_{ij}
\end{equation}
for normalised ${\cal{Y}}_{ij}$, subject to the constraints:
\begin{gather}
\label{eq:constraint-emplyment}
\sum_j {\cal{Y}}_{ij} = Y^{out}_i ,\\
\label{eq:constraint-attractiveness}
\sum_i\sum_j {\cal{Y}}_{ij} A_j = A^{tot} ,\\
\label{eq:constraint-ttw-cost}
\sum_i\sum_j {\cal{Y}}_{ij} C_{ij} = C^{tot} .
\end{gather}
The constraints in~\eqref{eq:constraint-emplyment} fix the total income flowing out of the employment area $i$ and towards all residence areas $j$.
The constraint in by~\eqref{eq:constraint-attractiveness} sets the total utility $A^{tot}$ that people obtain by living in areas $j$ with attractiveness $A_j$.
Our assumption is that people prefer to live in areas that are more populated, unless the population exceeds a saturation limit.
We also assume that people prefer areas where more services are available.
Therefore we define the attractiveness of a residence areas as $A_j = \log(f(P_j)\ S_j)$, where $f(P_j)$ is a function that assigns a score based on the population.
The population score $f(P_j)$ increases linearly with the population of $j$, until it reaches a point of saturation, after which additional population makes the score decrease.
Finally, constraint~\eqref{eq:constraint-ttw-cost} sets the total cost $C^{tot}$ of commuting between employment and residence areas.

The maximum entropy solution to this problem is 
\begin{equation}
\label{eq:max-ent-solution}
{\cal{Y}}_{ij} = \frac{Y_i^* e^{\alpha A_j - \gamma C_{ij}}}{Z_i} ,
\end{equation}
where $Z_i =  \sum_{j} e^{\alpha A_j - \gamma C_{ij}}$ are balancing factors.
The parameters $\alpha$ and $\gamma$ are the Lagrangian multipliers corresponding to the constraints in~\eqref{eq:constraint-attractiveness} and~\eqref{eq:constraint-ttw-cost}, and representing social disposition and impedance to travel, respectively.

We calibrate our model by identifying the optimal values $\hat{\alpha}$ and $\hat{\gamma}$ that agree with the initial output ${\cal{Y}}^0_{ij}$ best matching the actual flow of income $Y_{ij}$ given by Census (see Supplemental Material, Sec. 2).
The evolution of the services is then modelled yielding a prediction ${\cal{Y}}^*_{ij}$ within Greater Sydney.

\subsection{The Lotka-Volterra component}

The Lotka-Volterra component of the model is given by the following dynamics for the services $S_j$ over time $t$:
\begin{equation}
\label{eq:dynamics}
\frac{d S_j}{d t} = \epsilon({\cal{Y}}_j^{in} - R_jP_j - KS_j) ,
\end{equation}
where ${\cal{Y}}_j^{in}=\sum_i{\cal{Y}}_{ij}$ is the total income flowing into the suburb $j$ from all employment areas $i$,  $\epsilon$ defines the size of the changes and $K$ is a conversion factor such that $KS_j$ is the cost of running services $S_j$.
According to~\eqref{eq:dynamics}, if the remaining income ${\cal{Y}}_j^{in} - R_jP_j$ (analogous to discretionary income, which also subtracts taxes) flowing into the suburb is sufficient to compensate for the running costs of the services $KS_j$, then the services will grow, otherwise they will decrease.
Since the attractiveness $A_j$ is defined in terms of the services $S_j$, the former quantity also evolves.

\subsection{Fisher information and thermodynamic efficiency}

Following a recently established relationship~\cite{crosato2018thermodynamics}, the rate of change of the thermodynamic work can be determined using the Fisher information (see Supplemental Material, Sec. 3):
\begin{equation}
\frac{d\langle\beta W_{gen}\rangle}{d\alpha'} = -\int^\alpha_{\alpha_0}F(\alpha')d\alpha' + c(\alpha^0) ,
\end{equation}
where the Fisher information was calculated over the parameter $\alpha$ (fixing the parameter $\gamma$) as
\begin{equation}
\label{eq:fisher-info}
F(\alpha) \equiv \sum_i\sum_j{{\cal{Y}}^*_{ij}} \left(\frac{d\log{\cal{Y}}^*_{ij}}{d\alpha}\right)^2 = \sum_i\sum_j\frac{1}{{\cal{Y}}^*_{ij}} \left(\frac{d{\cal{Y}}^*_{ij}}{d\alpha}\right)^2
\end{equation}
for the maximum entropy solution ${\cal{Y}}^*_{ij}$.

Finally, we define the thermodynamic efficiency of urban transformation, for a given value of $\alpha$, as the reduction of entropy from the expenditure of work:
\begin{equation}
\label{eq:efficiency}
\eta \equiv \frac{-dH({\cal{Y}}_{ij}) / d\alpha}{d\langle\beta W_{gen}\rangle / d\alpha} .
\end{equation}
This quantity corresponds to a change $d\alpha$ and hence relates to a \emph{transformation}.
This approach is motivated by the notion of thermodynamic efficiency of computation~\cite{crosato2018thermodynamics}.


\section{Results}

\subsection{Abrupt urban transformations}

We explore the model predictions ${\cal{Y}}^*_{ij}$ over a range of values of the control parameters around their optimal values $\hat{\alpha}$ and $\hat{\gamma}$.
We then compute the entropy $H({\cal{Y}}^*_{ij})$ for the considered points within the phase diagram, tracing how the income distribution changes with respect to the control parameters.
Crucially, we observe that, while the entropy varies mostly linearly with respect to $\gamma$, it changes much more abruptly with the changes in $\alpha$ (see Supplemental Material, Sec. 4), indicating a phase transition.

\begin{figure}[b!]
\centering
\includegraphics[width=0.49\columnwidth]{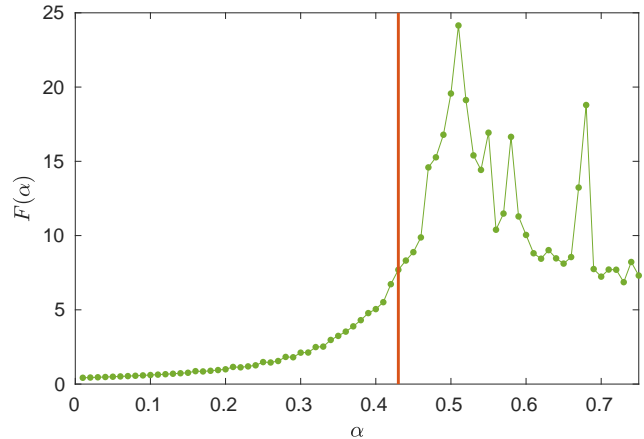}
\caption{Fisher information over $\alpha$, ranging from $0.001$ to $0.751$ with steps of $0.01$, with $\hat{\gamma}=0.15$ (best match).
The horizontal axis represents the values of $\alpha$, while the vertical axis represents the Fisher information of the flow or income ${\cal{Y}}^*_{ij}$.
The red vertical bar indicates the value $\hat{\alpha}=0.43$ for which ${\cal{Y}}^0_{ij}$ best matches the Sydney-2011 Census data, while the peak is at $\tilde{\alpha}=0.51$.
Note the second local peak at $\alpha=0.68$.
The maps on the sides of the plot are examples of Greater Sydney in the sprawling (left) and polycentric (right) phases.
}
\label{fig:fisher}
\end{figure}

However, in order to rigorously localise the abrupt change in the dynamics of income flow with respect to $\alpha$, we fix $\gamma$ at the optimal value $\hat{\gamma}$ and compute the Fisher information over the phase space of $\alpha$.
The result is shown in Fig.~\ref{fig:fisher}, which shows that the Fisher information peaks at $\tilde{\alpha}=0.51$.
This indicates that there is indeed a second-order phase transition in the space of $\alpha$, the critical point $\tilde{\alpha}$ of which is identified by the maximum value of the Fisher information, in line with the approach established in~\cite{crooks2007measuring, prokopenko2011relating, crosato2018thermodynamics}.
Fig.~\ref{fig:fisher} also shows $\hat{\alpha}=0.43$ that best matches Sydney-2011 Census data, which is lower than the critical value $\tilde{\alpha}$ but nevertheless is in the proximity of the phase transition, being located in the region where the Fisher information undergoes a rapid growth.

\begin{figure*}[t!]
\centering
\begin{tabular}{@{}m{0.925\textwidth}m{0.09\textwidth}@{}}
\includegraphics[width=0.4645\textwidth]{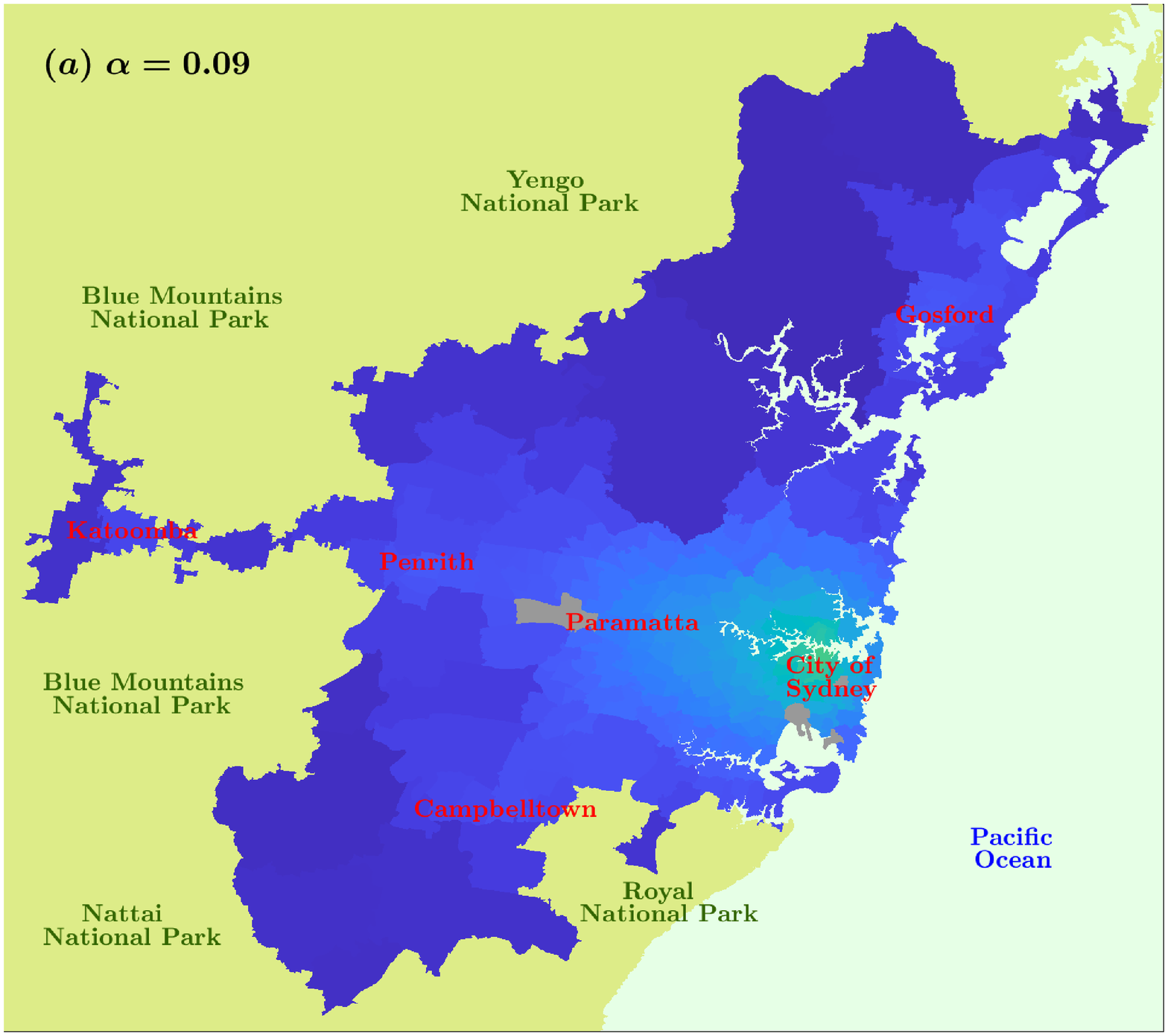}
\includegraphics[width=0.4645\textwidth]{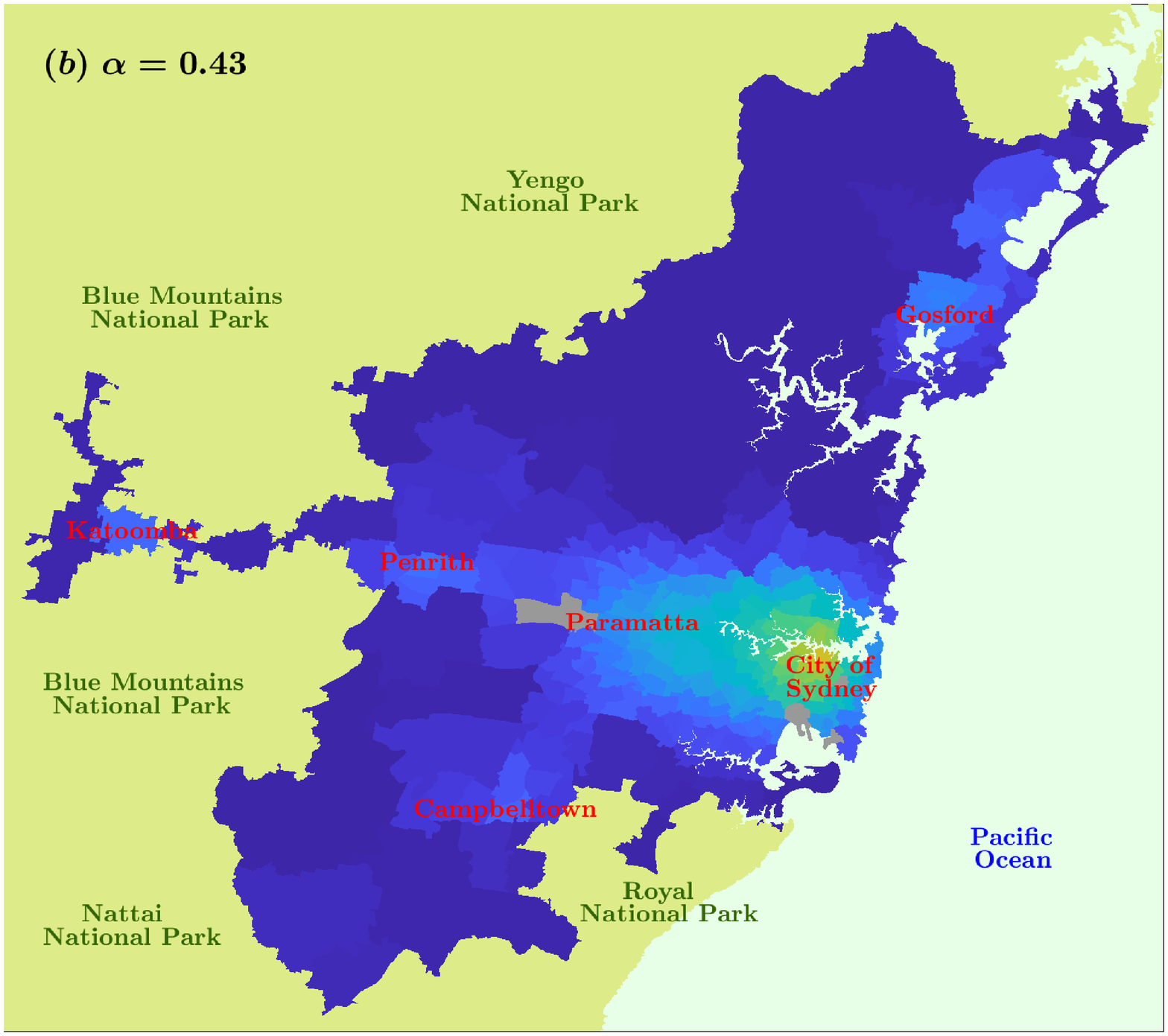}
\includegraphics[width=0.4645\textwidth]{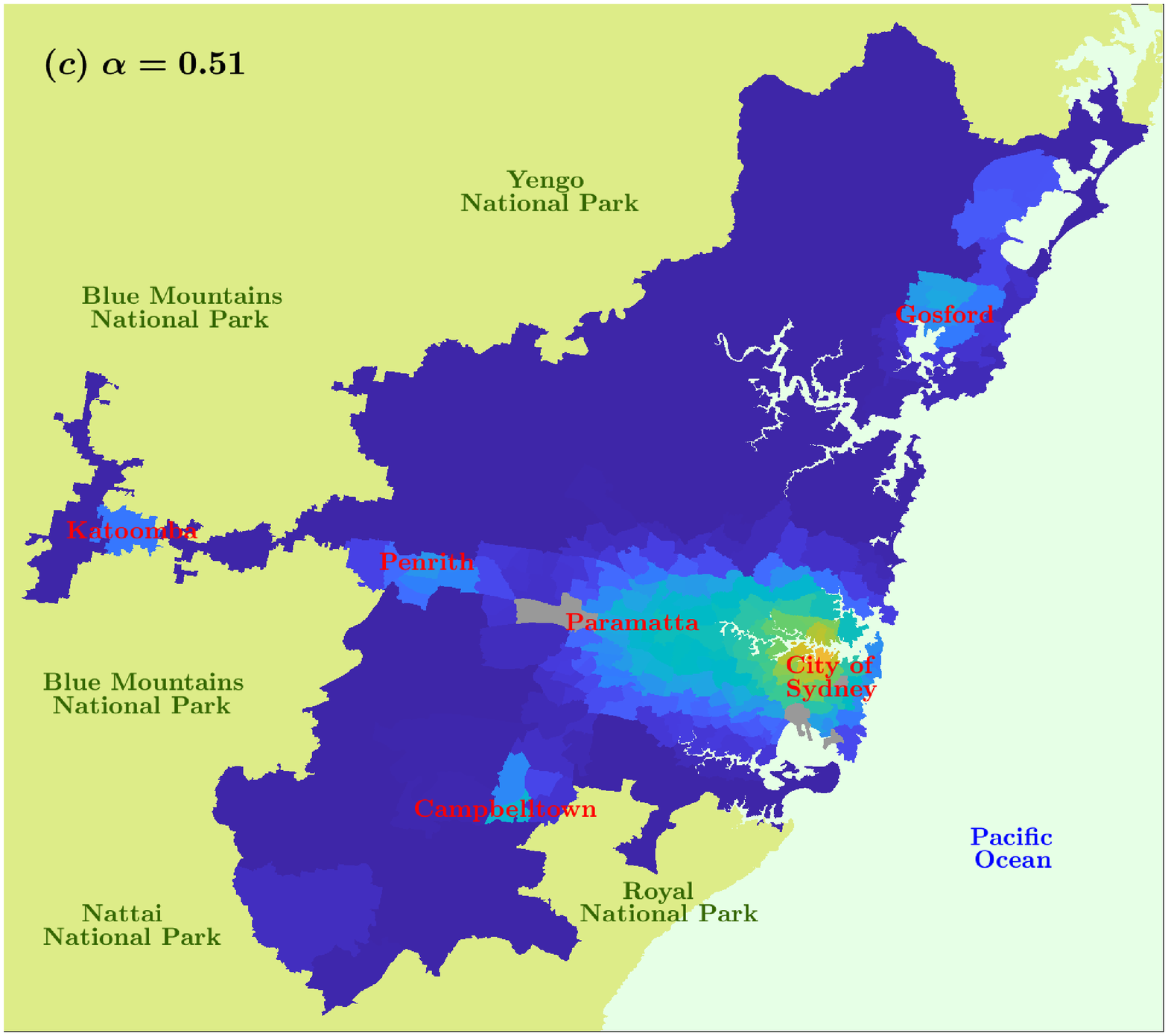}
\includegraphics[width=0.4645\textwidth]{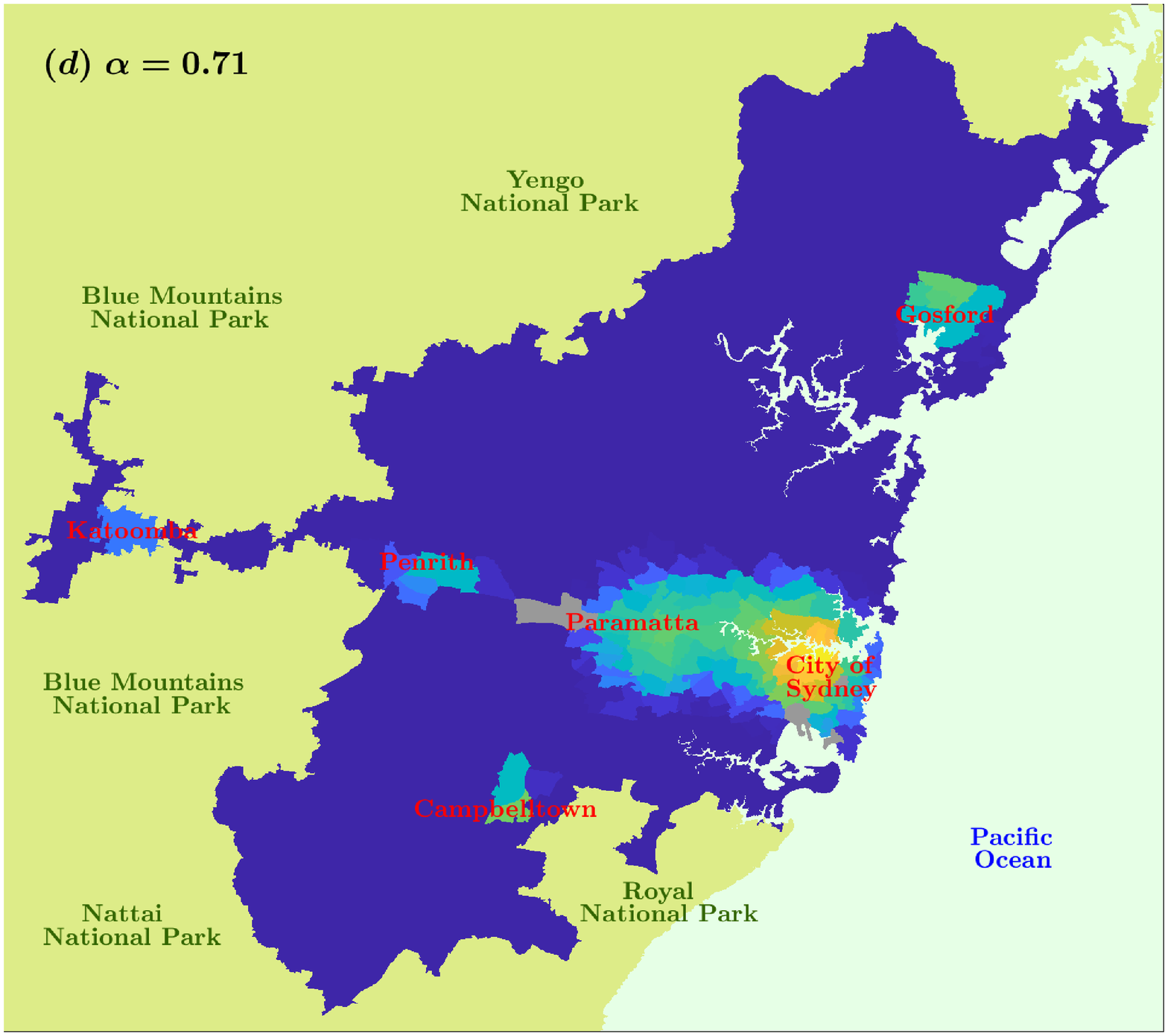}&
\includegraphics[width=0.065\textwidth]{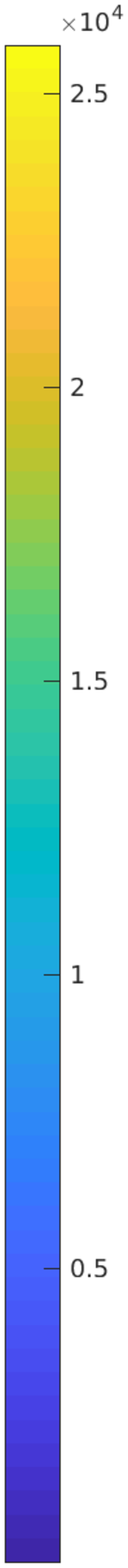}
\end{tabular}
\caption{The predicted population of Greater Sydney.
The region is partitioned into the 270 residence areas, which are coloured based on their population.
The grey areas represent national reserves and parks, Kingsford Smith airport and Port Botany which are not considered as residence areas.
(a) Predicted population with $\alpha=0.09$, corresponding to the sprawling urban phase.
(b) Predicted population with $\hat{\alpha}=0.43$, corresponding to the best match with Sydney-2011 Census data.
(c) Predicted population with $\tilde{\alpha}=0.51$, corresponding to the critical regime.
(d) Predicted population with $\alpha=0.71$, corresponding to the polycentric urban phase.
}
\label{fig:maps}
\end{figure*}

These results show that changes in the social disposition, away from its current value $\hat{\alpha}$, would significantly and abruptly change the flow distribution of income within Greater Sydney.
This has an immediate effect on the spatial distribution of the population, driving an urban transformation from the \emph{sprawling} phase to the \emph{polycentric} phase.
Fig.~\ref{fig:maps} shows the predicted population of Greater Sydney at fixed $\hat{\gamma}$ and four different values of $\alpha$: (a) a low value, far before the critical point, (b) the best match with the Sydney-2011 Census data $\hat{\alpha}$, (c) the critical point $\tilde{\alpha}$, and (d) a high value beyond the critical point.
Since the average income $I_i$ in the employment areas $i$ does not change, the population of each suburb $j$ is directly obtained from the flow of income ${\cal{Y}}^*_{ij}$ predicted by the model: $P_j=\sum_i({\cal{Y}}^*_{ij}/I_i)$.

For the low value of $\alpha$ (Fig.~\ref{fig:maps}(a)), corresponding to the sprawling urban phase, the model shows a quite homogeneous distribution of the population, with the areas around the City of Sydney and other major urban areas being only slightly more populated than the other surrounding areas.
As we move to $\hat{\alpha}$ (Fig.~\ref{fig:maps}(b)), the population aggregates around the major urban areas, although the City of Sydney seems to be the only highly populated area.
We note that this is the predicted population of Greater Sydney corresponding to the actual value of social disposition matching the Census data.
At the critical point $\tilde{\alpha}$ (Fig.~\ref{fig:maps}(c)) all the major urban areas become clearly more highly populated than the surrounding areas, and Greater Sydney starts to display a polycentric aggregation.
Finally, this polycentric aggregation becomes more pronounced in the polycentric urban phase, represented by the high value of $\alpha$ (Fig.~\ref{fig:maps}(d)): the areas of the City of Sydney, Parramatta, Penrith, Campbelltown and Gosford are clearly identifiable by a higher population compared to the surrounding.

Interestingly, Sydney-2011 profile, lying within the sprawling phase but near the phase transition, displays features of a polycentric metropolis, which accentuate beyond the critical point.
However, the dynamics of the polycentric phase are not steady (cf. Fig.~\ref{fig:fisher} for $\alpha > \tilde{\alpha}$), and so the transformations may suffer from tangible fluctuations and loss of predictability in social dynamics.
In fact, a secondary transformation is captured by the secondary local peak of the Fisher information, around $\alpha=0.68$ (see Fig.~\ref{fig:fisher}) and corresponding to noticeable population decline in the suburbs north of Gosford (see Supplemental Movie S1).
``Double percolation'' phase transitions have been observed in clustered complex networks with spatiotemporal dynamics~\cite{colomer2014double}, and multiple peaks detected by the Fisher information may relate to this phenomenon, given the clustered connectivity of urban aggregations.

\subsection{Deepening the thermodynamic analogy}

An important consideration in making a rigorous thermodynamic analogy is a choice of the protocol according to which the control parameters are varied, so that the corresponding changes in the required work, energy and configuration entropy, as well as symmetry breaking~\cite{nikoghosyan2016universality}, can be traced.
Specifically, we consider a quasi-static protocol varying $\alpha$, at the expenditure of some required work, and driving changes from the sprawling urban phase to the polycentric phase, across the phase transition.
For a quasi-static protocol the required work is minimal, i.e., the work matches the free energy of the system.

\begin{figure}[t!]
\centering
\subfigure[]{\includegraphics[width=0.47\columnwidth]{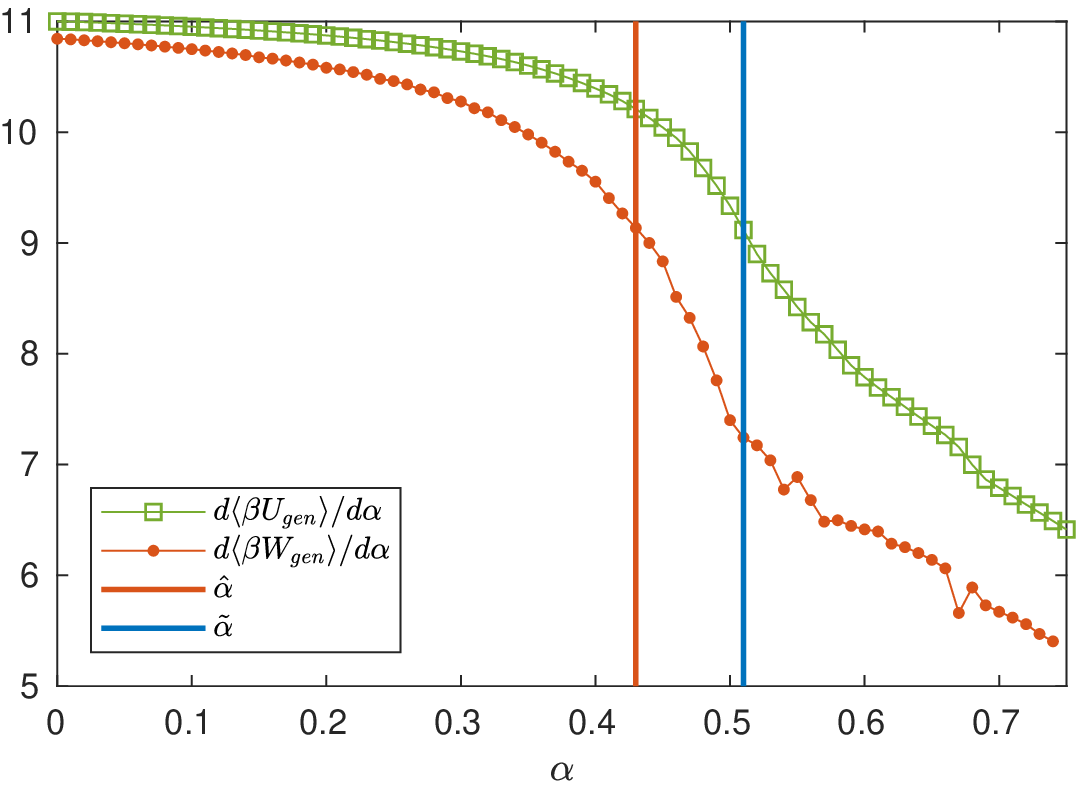}}
\subfigure[]{\includegraphics[width=0.51\columnwidth]{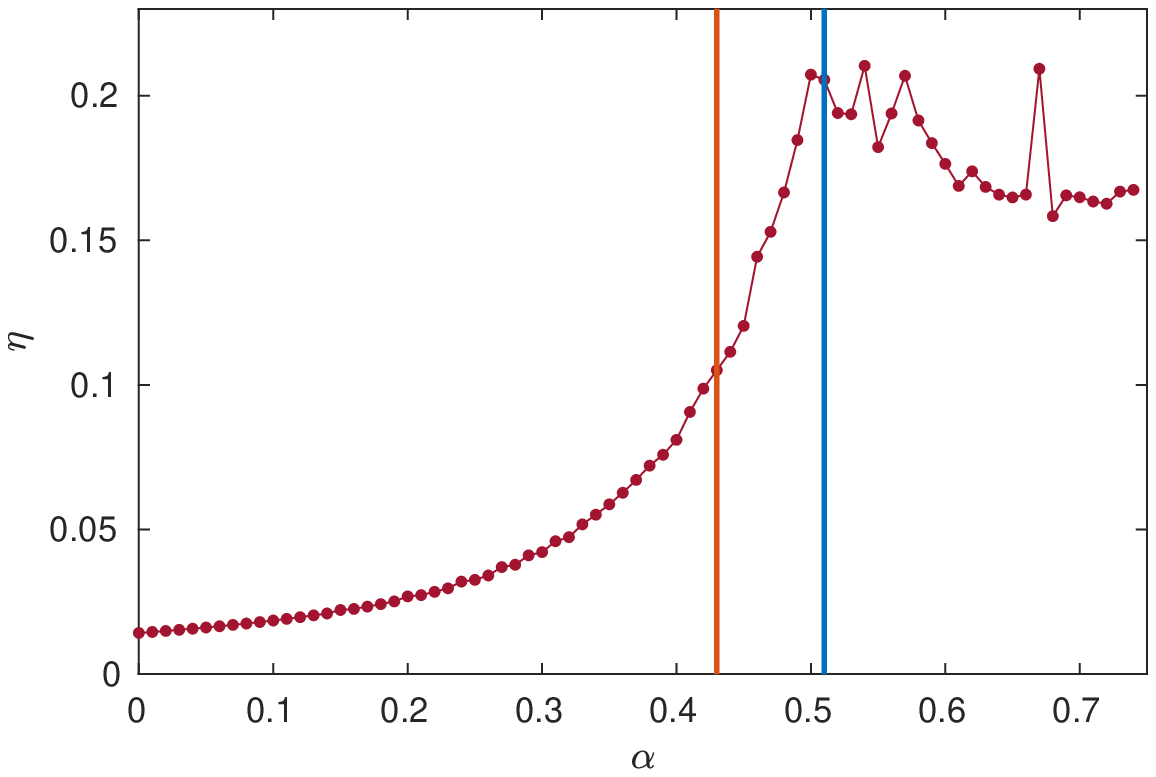}}
\caption{(a) The rates of change of the work $W_{gen}$ (green line) and the internal energy $U_{gen}$ (red line).
(b) The thermodynamic efficiency of urban transformation, $\eta$, defined as the order-to-work ratio.
In both graphs the vertical lines represent the values of the control parameter $\hat{\alpha}$ (red line), which best matches the Sydney-2011 Census data, and $\tilde{\alpha}$ (blue line), the critical value.}
\label{fig:thermo}
\end{figure}

It has been recently shown that for quasi-static processes the second derivative of the generalised work $W_{gen}$ with respect to a control parameter is proportional to the negative of the Fisher information~\cite{crosato2018thermodynamics}.
We refer to generalised work in the sense of Jaynes~\cite{jaynes1957information} (for more details about generalised quantities and their relationship with the Fisher information see Supplemental Material, Sec. 3).
Given this relationship, we obtain the rate of change of the work with respect to $\alpha$ by numerically integrating the negative of the Fisher information in Fig.~\ref{fig:fisher}.
The result is shown in Fig.~\ref{fig:thermo}(a), demonstrating that the rate of change of the work decreases with $\alpha$, with this change becoming more pronounced in the proximity of the optimal value $\hat{\alpha}$, being steepest around the critical point $\tilde{\alpha}$.

Fig.~\ref{fig:thermo}(a) also shows the rate of change of the internal energy of the system $U_{gen}$.
This quantity is obtained from the rates of change of the work $W_{gen}$ and the configuration entropy 
$H({\cal{Y}}_{ij})$---according to the first law of thermodynamics (in the case of quasi-static processes) a change in the internal energy corresponds to the sum of the changes in entropy and work: $\Delta\langle U_{gen}\rangle = \Delta\langle W_{gen}\rangle + \Delta H({\cal{Y}}_{ij})$, where the angle brackets represent average values over the ensemble.
The rate of change of the internal energy decreases with $\alpha$ similarly to the rate of change of the work.
The difference between the two rates of change (i.e., the rate of change of the entropy) is larger around the critical point, when the flow of income exhibits a tendency towards the polycentric phase.

The thermodynamic efficiency of urban transformations for Greater Sydney, $\eta$, is shown in Fig.~\ref{fig:thermo}(b).
It can be seen that $\eta$ is very low in the sprawling phase, increases towards the phase transition and then tends to slowly decrease, while also exhibiting the secondary local peak.
Interestingly, this ratio is in a midrange for the value $\hat{\alpha}$ corresponding to Sydney-2011 Census data.
It is also evident that the social disposition estimated from the Sydney-2011 data characterises the sprawling phase, distinct from the polycentric phase.


\section{Discussion}

The transition of cities between different patterns of urban settlement (dispersed, monocentric, polycentric, etc.) has become a central problem in urban planning.
In this study we investigated the urban dynamics from a statistical mechanical viewpoint, deriving a thermodynamic description and applying it to a case study of Greater Sydney.
This approach complements the maximum entropy principle with dynamics of evolving urban structures at different time scales, identifies phase transitions using Fisher information and quantifies the thermodynamic efficiency of urban transformations.

The model has been calibrated to Census data and geospatial datasets and exhibits a clear phase transition between a dispersed configuration, in which the population settles homogeneously within Greater Sydney, and a polycentric configuration, in which the population aggregates in a few highly populated urban clusters.
Two salient quantities are represented by the attractiveness of suburbs, in terms of services available to the population, and the commuting costs.
The phase transition was shown to be induced by the control parameter accounting for social disposition---a factor balancing the suburbs' attractiveness---rather than the parameter tracking travel impedance.

A recent plan by the Greater Sydney Commission~\cite{greater2017greater} envisaged a tripartite Greater Sydney region, with a western parkland city, a central river city around greater Parramatta, and an eastern harbour city.
As shown in our study, such a tripartite arrangement is possible only under a narrow set of constraints and importantly lies in the polycentric urban phase, separated from the current sprawling phase by a phase transition.
Thus, a major urban transformation will inevitably pass through a critical regime with its inherent fluctuations and loss of predictability in social dynamics.
Nevertheless, a set of policies informed by a quantitative approach may steer this transformation exploiting the resultant gain in efficiency.
A wide class of other urban scenarios may also be considered within the proposed approach, in which the concise thermodynamic descriptions are derived purely based on probability distributions estimated from available data.


\appendix

\section{Greater Sydney and data sources}

Greater Sydney is an urban area covering more that \num{12000} square kilometres, delimited in all directions by either the Pacific Ocean or by the several surrounding national parks.
It includes the City of Sydney as well as other urban agglomerations such as Parramatta, Penrith, Campbelltown and Gosford, for a total population of approximately 5 million.
According to 2011 Census data, the working population of Greater Sydney is 1.8 million.
People daily commute between residence areas (or suburbs), where they live, and employment areas, where they work.
The territory is partitioned into \num{270} residence areas and \num{2156} employment areas.
The data used for this study was provided by the Australian Bureau of Statistics.
This includes the geospatial data of the areas of employment and residence, as well as the Census data for year 2011.

The employment areas are defined by the standard Destination Zone (DZN), which was designed by the New South Wales transport authority in order to spatially classify employment places, with the purpose of analysing commuting data and developing transport policies.
The standard Statistical Area Level 2 (SA2), as defined by the Australian Statistical Geography Standard, was used for the residence areas.
Geographical areas of level SA2 represent small communities that closely interact socially and economically.
The population of these areas can vary from \num{3000} to \num{25000} individuals, with an average population of \num{10000} individuals.

The Census data for year 2011 was geographically classified by the Australian Bureau of Statistics in accordance with the both geographical areas DZN and SA2, and included the travel-to-work matrix $T_{ij}$, the average weekly income $I_i$ and the average weekly rent $R_j$, for all DZN areas $i$ and SA2 areas $j$.
The Census data also included the amount of people who work in food retailing stores (including supermarkets, grocery stores, meat and fish stores, fruit and vegetables stores and liquor stores) that are located in specific SA2 areas.
This data was utilised to estimate the amount of goods, matching the services $S_j$ available in each residence area.

\begin{figure}[b!]
\centering
\includegraphics[width=0.49\columnwidth]{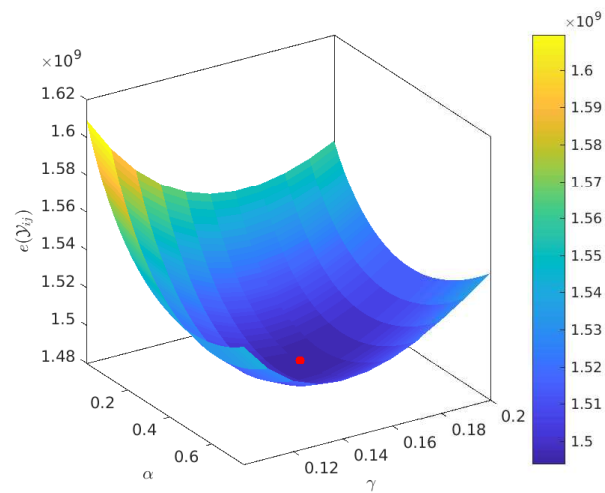}
\caption{Calibration of the parameters $\alpha$ and $\gamma$.
The horizontal axes represent values of $\alpha$ and $\gamma$ within the considered ranges, while the vertical axis represents the difference between the income flow ${\cal{Y}}^0_{ij}$ produced by the model and the actual income flow $Y_{ij}$ given by Sydney-2011 Census data, calculated as  $e({\cal{Y}}^0_{ij}) = \sum_i\sum_j \big|{\cal{Y}}^0_{ij}-Y{ij}\big|$.
The red dot represents the optimal values $\hat{\alpha}$ and $\hat{\gamma}$.}
\label{fig:error}
\end{figure}

The cost of travelling $C_{ij}$ was estimated as the Euclidean distance between the centres of the employment and residence areas.
An alternative approach would be to calculate the time of travelling using Google Maps or OpenStreetMap data (however, that requires access to high resolution data which is not immediately available at the required spatial scales).

\section{Calibration of the model}

The model was calibrated by identifying the optimal values $\hat{\alpha}$ and $\hat{\gamma}$ for which the output ${\cal{Y}}^0_{ij}$ best matches the actual flow of income $Y_{ij}$ of Sydney-2011.
The difference between actual and predicted flow of income was estimated as the sum of (the absolute values of) the differences between all values of the matrices ${\cal{Y}}^0_{ij}$ and $Y_{ij}$, which is $e({\cal{Y}}^0_{ij}) = \sum_i\sum_j \big|{\cal{Y}}^0_{ij}-Y{ij}\big|$.
The result is shown in Fig.~\ref{fig:error}.

\section{Thermodynamic analysis}

Let us consider the state functions $X_m(x)$ that describe a physical system over its configurations $x$.
In a stationary state, the Gibbs measure defines the probability of the states of the system:
\begin{equation}
\label{eq:gibbs-measure}
p(x|\theta) = \frac{1}{Z(\theta)}e^{-\beta H(x,\theta)} = \frac{1}{Z(\theta)}e^{-\sum_m \theta_m X_m(x)} ,
\end{equation}
where $\theta_m$ are thermodynamic variables, $\beta=1/k_bT$ is the inverse temperature $T$ ($k_b$ is the Boltzmann constant), $H(x,\theta)$ is the Hamiltonian defining the total energy at state $x$, and $Z(\theta)$ is the partition function~\cite{brody1995geometrical, crooks2007measuring}.
The Gibbs free energy of such system is:
\begin{equation}
\label{eq:gibbs-potential}
G(T,\theta_m) = U(S,\phi_m) - TS - \phi_m\theta_m ,
\end{equation}
where $U$ is the internal energy of the system, $S$ is the configuration entropy and $\phi_m$ is an order parameter.
Let us also consider the generalised internal energy $U_{gen}$ in the sense of Jaynes~\cite{jaynes1957information}, such that
\begin{equation}
\langle\beta U_{gen}\rangle = U(S,\phi_m) - \phi_m\theta_m ,
\end{equation}
where the angle brackets represent average values over the ensemble.
The generalised first law holds $\langle\beta U_{gen}\rangle = \langle\beta Q_{gen}\rangle + \langle\beta W_{gen}\rangle$, where $Q_{gen}$ and $W_{gen}$ are, respectively, the generalised heat and the generalised work.

The Fisher information~\cite{fisher1922mathematical} measures the amount of information that an observable random variable $X$ carries about an unknown parameters $\theta = [\theta_1,\theta_2,\dots,\theta_M]^T$.
If $p(x|\theta)$ is the probability of the realisation $x$ of $X$ given the parameters $\theta$, the Fisher information matrix is defined as
\begin{equation}
\label{eq:fisher-matrix}
F_{mn}(\theta) = E \Bigg[ \bigg( \frac{\partial \ln p(x|\theta)}{\partial\theta_m} \bigg) \bigg( \frac{\partial \ln p(x|\theta)}{\partial\theta_n} \bigg) \Bigg| \theta \Bigg] ,
\end{equation}
where the function $E(y)$ is the expected value of $y$.
For a physical system described by the Gibbs measure in \eqref{eq:gibbs-measure}, the Fisher information has several physical interpretations, e.g., it is equivalent to the thermodynamic metric tensor $g_{mn}(\theta)$, is proportional to the second derivative of the free entropy $\psi = \ln Z = -\beta G$, and to the derivatives of the corresponding order parameters with respect to the collective variables~\cite{brody1995geometrical, brody2003information, janke2004information, crooks2007measuring, prokopenko2011relating}:
\begin{equation}
 F_{mn}(\theta) = g_{mn}(\theta) = \frac{\partial^2\psi}{\partial\theta_m\partial\theta_n}  = \beta\frac{\partial\phi_m}{\partial\theta_n}.
\end{equation}
Furthermore~\cite{crosato2018thermodynamics},
\begin{equation}
\label{eq:fisher-diff}
 F(\theta) = \frac{d^2S}{d\theta^2}- \frac{d^2\langle\beta U_{gen}\rangle}{d\theta^2}.
\end{equation}

Under a quasi-static protocol the total entropy production is zero, and therefore any change in the configuration entropy due to the driving process is matched by the flow of heat to the environment:
\begin{equation}
\label{eq:entropy-heat-qs}
\frac{dS}{d\theta} = \frac{d \langle\beta Q_{gen}\rangle}{d\theta} .
\end{equation}
Thus, combining~\eqref{eq:fisher-diff} and~\eqref{eq:entropy-heat-qs} with the first law of thermodynamics yields another important result for the generalised work  $W_{gen}$~\cite{crosato2018thermodynamics}:
\begin{equation}
\label{eq:fisher-work-curvature}
F(\theta) = - \frac{d^2\langle\beta W_{gen}\rangle}{d\theta^2} .
\end{equation}

\section{Entropy and a proxy of order parameter}

\begin{figure}[b!]
\centering
\includegraphics[width=0.49\columnwidth]{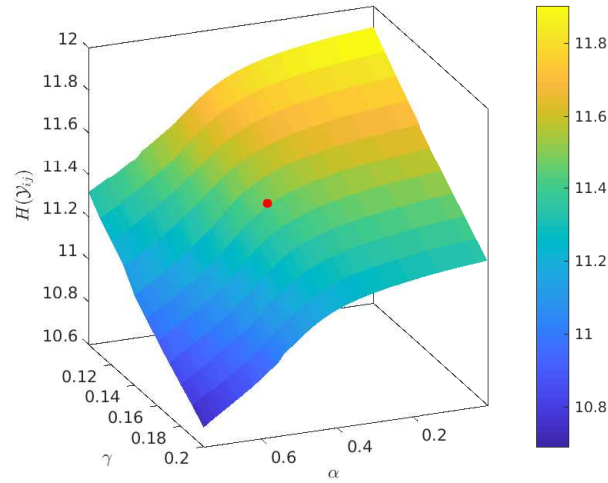}
\caption{Entropy of ${\cal{Y}}^*_{ij}$ after the services $S_j$ have evolved to reach an equilibrium.
The horizontal axes represent values of $\alpha$ and $\gamma$ within the considered ranges, while the vertical axis represents the entropy $H({\cal{Y}}_{ij})$ at corresponding values of $\alpha$ and $\gamma$.
The red dot indicates the combination of $\hat{\alpha}$ and $\hat{\gamma}$ for which ${\cal{Y}}^0_{ij}$ best matches Sydney-2011 Census data.}
\label{fig:entropy}
\end{figure}

A higher entropy indicates a more homogeneous distribution of the income to all suburbs, while a lower entropy indicates a less balanced distribution of the income biased towards one or few suburbs.
We observe that the entropy decreases with both parameters $\alpha$ and $\gamma$ (see Fig.~\ref{fig:entropy}).
This behaviour is expected and has a clear interpretation.
If the social disposition $\alpha$ is low, people have modest preference for attractive suburbs and thus settle (and move their income) more homogeneously within the region, while if $\alpha$ is high people tend to aggregate around the areas with the higher attractiveness.
Similarly, if travel impedance $\gamma$ is low people are less concerned about high travel costs, and therefore can settle at any distance from their work place, while if $\gamma$ is high people prefer to live closer to their employment areas to incur lower commuting costs.

\begin{figure}[b!]
\centering
\includegraphics[width=0.49\columnwidth]{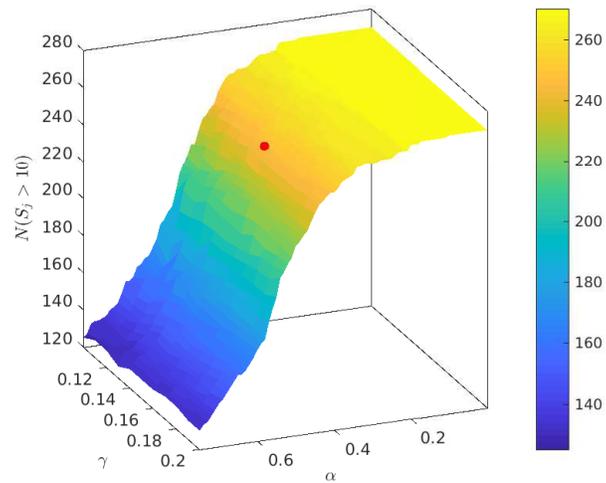}
\caption{The number of suburbs $N(S_j>10)$ with more than $10$ units of services available, after the services $S_j$ have evolved to reach an equilibrium.
The horizontal axes represent values of $\alpha$ and $\gamma$ within the considered ranges, while the vertical axis represents $N(S_j>10)$ at corresponding values of $\alpha$ and $\gamma$.
The red dot indicates the combination of $\hat{\alpha}$ and $\hat{\gamma}$ for which ${\cal{Y}}^0_{ij}$ best matches Sydney-2011 Census data.}
\label{fig:services}
\end{figure}

To formalise this intuition, one typically introduces and traces corresponding order parameters.
This is however hindered by an incomplete statistical-mechanical description of the system, and we first illustrate a simpler approach which considers a proxy of an order parameter.
Such a proxy characterises the equilibrium distribution of the services $S_j$, for different values of $\alpha$ and $\gamma$, in terms of the number of suburbs in which the amount of available services exceeds a threshold, i.e., ``services-abundant'' suburbs.
Fig.~\ref{fig:services} shows the number of services-abundant suburbs, that is $N(S_j>10)$, for different values of $\alpha$ and $\gamma$, after the urban evolution has converged.
Again, as with the entropy dynamics, the variation of $\gamma$ does not greatly affect the number of services-abundant suburbs.
On the contrary, this number displays an abrupt change with respect to $\alpha$: for low values of the social disposition all 270 suburbs are services-abundant, but as $\alpha$ increases the number of services-abundant suburbs reduces quickly past a specific value of $\alpha$.
At high values of social disposition approximately 120 residence areas remain services-abundant.

Fig.~\ref{fig:entropy} and Fig.~\ref{fig:services} also show the values $\hat{\alpha}$ and $\hat{\gamma}$ which best matches Sydney-2011 Census data (the red dot on either the entropy or the $N(S_j>10)$ surfaces).
This value is within a close proximity to the social disposition where the abrupt change is observed.

\bibliographystyle{vancouver}

\section*{Acknowledgements}
All authors were supported by The University of Sydney's DVC Research Strategic Research Excellence Initiative (SREI-2020) project, ``CRISIS: Crisis Response in Interdependent Social-Infrastructure Systems'' (Grant No. IRMA 194163).
E.C. was supported by the University of Sydney's ``Post-graduate Scholarship in the Field of Complex Systems'' from Faculty of Engineering \& IT and by a CSIRO top-up scholarship.
M.P. was supported by the ARC Discovery project ``Australian housing market risks: simulation, modelling and analysis'' (Grant No. DP170102927).
Sydney Informatics Hub at the University of Sydney provided access to HPC computational resources that have contributed to the research results reported within the paper.

\end{document}